\documentclass[reqno,12pt]{amsart}
\usepackage{amsfonts}
\usepackage{amssymb}
\usepackage{graphicx}
\usepackage{fullpage}
\usepackage{color}
\usepackage{url}
\usepackage{epstopdf}

\def\R{\hbox{{\rm I}\kern-0.2em{\rm R}\kern0.2em}}%mathematical R for reals
\def\D{\hbox{{\rm I}\kern-0.2em{\rm D}\kern0.2em}}

\def\be{\begin{equation}}
\def\ee{\end{equation}}

\def\({\left(}
\def\){\right)}
\def\[{\left[}
\def\]{\right]}
\def\bc{\begin{center}}
\def\ec{\end{center}}

\begin{document}

%\title{A study of positive energy condition of new solutions obtained via symmetries of the geodesic Lagrangian in Bianchi V spacetimes}
%\title{A study of positive energy condition of new solutions in Bianchi V spacetimes}
\title{A study of positive energy condition in Bianchi V spacetimes via Noether symmetries}

\author{Sajid Ali, Ibrar Hussain}
\address{Dept. of Basic Sciences, School of Electrical Engineering and Computer Sciences, National University of Science and Technology, Campus H-12, Islamabad 44000, Pakistan.}
\email{sajid{\_}ali@mail.com; ibrar.hussain@seecs.nust.edu.pk}
\maketitle
\begin{abstract}
In this paper we use Noether symmetries of the geodesic Lagrangian in Bianchi V spacetimes to study various cosmological solutions of Einstein's field equations. Our first result is the identification of the subalgebras of Noether symmetries of the equations of motion in such spacetimes with dimension 4, 5, 6, 7, 9 or 10 of the maximal algebra of Lie point symmetries of dimension 13. Secondly we give physical interpretation of new cosmological solutions which satisfy positive energy condition and yield critical bounds on the expansion coefficient $\alpha$, in which the underlying non-flat spacetimes carry interesting physical properties. Specifically the energy density behaves in one of the following ways. (i) It is positive and constant for all time. (ii) It varies with time and attains a global maximum after some time and then asymptotically converges to zero. (iii) It increases for all time and attains a maximum value at the asymptotic limit $t\rightarrow \infty$. In particular a non-flat spacetime is obtained that mimics the expansion in a flat FRW universe dominated by vacuum energy such that the expansion factor has the same form in both. However, the energy density is dynamical in the former.
\end{abstract}

% In this paper we investigate Noether symmetries in Bianchi type V spacetimes. Our study is mainly divided into three  parts. (a) First we carry out complete classification of such spacetimes using Lie algebras of their Noether symmetries and establish that the dimension of Noether algebra is 4, 5, 6, 7, 9 or 10. (b) Then we employ Noether's theorem to obtain first integrals in each case and discuss their physical consequences on geodesic motion. (c) Lastly we give a brief comparison of Noether symmetry algebras, Lie symmetry algebras with the algebras of conformal Killing vectors, curvature and Weyl collineations.\\

%%%%%%%%%%%%%%%%%%%%%%%%%%%%%%%%%%%%%%%%%%%%%%%%%%%%%%%%%%%%%%%%%%%%%%%%%%%%%%%%%%%%%%%%%%%%%%%%%%%%%%%%%%%%%%%%%%%%%%%
%%%%%%%%%%%%%%%%%%%%%%%%%%%%%%%%%%%%%%%%%%%%%%%%%%%%%%%%%%%%%%%%%%%%%%%%%%%%%%%%%%%%%%%%%%%%%%%%%%%%%%%%%%%%%%%%%%%%%%%
%%%%%%%%%%%%%%%%%%%%%%%%%%%%%%%%%%%%%%%%%%%%%%%%%%%%%%%%%%%%%%%%%%%%%%%%%%%%%%%%%%%%%%%%%%%%%%%%%%%%%%%%%%%%%%%%%%%%%%%
%%%%%%%%%%%%%%%%%%%%%%%%%%%%%%%%%%%%%%%%%%%%%%%%%%%%%%%%%%%%%%%%%%%%%%%%%%%%%%%%%%%%%%%%%%%%%%%%%%%%%%%%%%%%%%%%%%%%%%%
\section{Introduction}
Finding exact solutions of the Einstein's field equations (EFEs) is one of the important and an old problem in General Relativity (GR). The	 search for new solutions opens new avenues to our understanding of the universe. These equations are the result of Einstein's revolutionary idea that the existence of matter induces curvature in a spacetime. This is responsible for an inward pull of all surrounding objects, thereby completely changing the standard perspective of gravitational force \cite{misner,inv}. Recent developments in astrophysics and cosmology reveal that our universe is mainly composed of a mysterious form of matter (dark energy and dark matter) which is the main cause of cosmic acceleration at a faster pace today \cite{per, rie}. This astonishing observation laid down the basis of numerous phenomenological models and led to the departure from the standard solutions of EFEs in the thrust of new exact solutions possessing revived geometrical and dynamical properties. Our attempt in this direction is to exploit the geometrical features in homogeneous cosmological solutions of these equations which mainly arise from Noether symmetries.

Suppose $M$ is a four-dimensional smooth manifold equipped with a non-degenerate metric $g$ of Lorentzian signature $(+,-,-,-)$. The EFEs of the universe in standard gravitational units $c=1=G$, are a system of partial differential equations $G_{\mu\nu} = T_{\mu\nu}$ ($\kappa =1$), where $G_{\mu\nu}$ is the Einstein tensor which contains all the basic geometric properties of spacetimes and $T_{\mu\nu}$ is the stress-energy tensor that describes the density and flux of energy and momentum in spacetimes \cite{inv}. The exact solutions of EFEs either arise from the geometrical consideration of spacetimes or by studying physical characteristics of matter or a combination of both. In cosmology, homogeneous spacetimes of dimension $1+3$ are characterized using Bianchi classification of real Lie algebras of dimension three which resulted into nine classes I, II, ...,IX \cite{kramer}. The famous cosmological solutions of EFEs known as Friedmann-Lemaître-Robertson-Walker (FRW) metrics are both isotropic and homogeneous and are particular cases of Bianchi types I, V, VII and IX \cite{misner}. Therefore it is important to investigate other Bianchi homogeneous cosmologies with regard to their invariant geometrical properties. Indeed the authors \cite{cot}, have done the symmetry analysis of all Bianchi spacetimes in the presence of a dynamical field. In \cite{cam}, authors have classified Bianchi V spacetimes using symmetries of the curvature and Weyl tensors known as curvature collineations (CC) and Weyl collineations (WC), respectively.

The concept of symmetries is closely related to the invariance of geometrical quantities under certain diffeomorphism on the manifold that could change the manifold however it keeps the underlying structure intact. This is a well known problem of equivalence of the two geometric objects under diffeomorphisms \cite{olver, karl}. The invariance of geometrical quantities which are described by tensor fields on the manifold under certain diffeomorphisms is ensured if the Lie derivative vanishes along the vector field representing a flow. Suppose $\bf{T}$ is a tensor field of any type on the manifold $M$. The key to check the invariance of certain geometrical quantities is to use the definition of Lie derivatives and require that
$\pounds_{\bf{X}} \bf{T} =0$, where $\pounds_{\bf{X}}$ is the Lie derivative along the vector field $\bf{X}$ \cite{kramer}. In case of the metric tensor the symmetries are known as conformal Killing vectors (CKV) provided $\pounds_{X} g_{\mu\nu} = 2 \Psi g_{\mu\nu}$, where $\Psi(x^{\mu})$ is known as a conformal factor. If $\Psi_{,\mu\nu}  \neq 0$, then a CKV is said to be proper otherwise it reduces to a special CKV if $\Psi_{,\mu\nu}=0$ and $\Psi_{,\mu} \neq 0$. The homotheties (HV) and Killing vectors (KV) arise if $\Psi_{,\mu} =0$ and $\Psi=0$, respectively. Furthermore the symmetries of the curvature tensor are obtained by replacing the tensor field with curvature tensor  $\pounds_{X}R^{\mu}_{~\nu\rho\sigma}=0$. In order to investigate pure gravitational field of spacetimes Weyl tensor plays a significant role as it is conformally invariant and its symmetries are obtained by the same condition that the Lie derivative of the Weyl tensor $C^{\mu}_{~\nu\rho \sigma}$ vanishes.

Noether symmetries play an essential role in finding conservation laws of the equations of motion with the use of Noether theorem \cite{olver,ov}. Sometimes it is difficult to obtain exact solutions of the equations of motion, however the dynamics can be reduced by investigating the invariant properties of the system provided the problem under consideration is variational and there exist a Lagrangian. Noether's theorem provides an explicit formula of a conserved quantity for each continuous symmetry transformation that leaves the action invariant \cite{noe}. In addition Noether symmetry analysis reduces the problem by specifying the unknown functions that appear in the Lagrangian. This line of approach has been followed by several authors, notably in the pioneering works of S. Capozziello \cite{cap1,cap}, G. E. Prince \cite{prince, prince1} and M. Tsamparlis \cite{tsam,tsam1}. The connection between symmetries of the underlying manifold and that of the differential equations was discussed in \cite{tooba}. The classification of spherically symmetric static spacetimes via Noether symmetries is done in \cite{far}. Besides approximate Lie symmetries was used to resolve the problem of energy in general relativity in \cite{ibr, ibr1}. In this paper our main interest is the classification of Bianchi V spacetimes using Noether symmetries. These symmetries provide crucial physical information about the conserved quantities of a physical system.

We start our investigation by considering the geodesic motion in Bianchi type V spacetimes. In such spacetimes there are three arbitrary functions whose specification arise from the presence of Noether symmetries. We give a complete classification of Noether algebras along with first integrals for each case. It is found that Noether algebra of the symmetries of the geodesic Lagrangian in such spacetimes can have dimension, $4-7$, $9$ or 10. The algbera of Lie point symmetries of the equations of motion is of dimension, $5-8$, 10, 12 or 13. The connection between Lie and Noether symmetries with symmetries of the spacetimes like HVs, projective collineations (PCs) and CCs has already been discussed in \cite{tooba, tsam}. We highlight the important features of new cosmological solutions in the light of above results.

It is well known that there are three cosmological models of our universe which are filled with vacuum energy, radiation and matter \cite{misner}. The large astrophysical data suggest a widely accepted view that our universe is nearly flat. It would be interesting to investigate as to what degree we can relax the condition of flatness of the underlying manifold such that the non-flat spacetime still carry all the important features of a realistic and viable cosmological model. It turns out that using Noether symmetries we can specify certain non-flat spacetimes which possess very nice physical properties. In particular we determine a non-flat spacetime that mimics the behavior of a flat spacetime filled with vacuum energy. Interestingly the expansion factor (so is the Hubble parameter) in the non-flat and flat spacetimes is same. The energy density is fixed in the flat model (filed with vacuum energy) and dynamical in the other case. Based on our finding we further investigate non-flat spacetimes and compare these with flat spacetimes dominated by matter (Einstein-de Sitter universe) or radiation.

The paper is subsequently divided into three main parts. The Noether symmetry analysis is presented in the next Section where we give preliminaries to the symmetry approach. In the third Section we investigate the physical characteristics of such spacetimes and examine the implications of positive energy condition on the solutions. To confront our results with other extended theories of gravity we consider a particular Bianchi V spacetime which meets the positive energy condition and modify our results in $f(R)-$gravity. It is found that $f(R)\propto R^{3/2}$, for a dust cloud in an anisotropic Bianchi V spacetime. Lastly we summarize our results in the last Section. 

\section{Noether symmetry Analysis}
The Bianchi type V spacetimes have the form
\begin{equation}
\mbox{d}s^2 = \mbox{d}t^2 - A(t)^2 \mbox{d}x^2 - e^{\alpha x} \left ( B(t)^2 \mbox{d}y^2 + C(t)^2 \mbox{d}z^2\right ) ,
\end{equation}
where $A,B,C,$ are three non-zero arbitrary functions of the cosmic time $t$. The expansion factor $\alpha$, is a non-zero constant which has the units of inverse length. The above spacetimes represent a specially homogeneous and anisotropic cosmologies as the coefficients depend on the time variable $t$. Since these are non-static spacetimes and thus do not admit a time like Killing vector field. All Binachi V spacetimes admit a three-dimensional Lie algebra of spacelike KVs (corresponding to the isometry group $G_{3}$, that acts transitively on the spacelike hypersurfaces) \cite{kramer}
\begin{align*}
 &
 X_{1} = \frac{\partial}{\partial y},\quad X_{2} = \frac{\partial}{\partial z}, \quad X_{3} = 2\frac{\partial}{\partial x}- \alpha y\frac{\partial}{\partial y}- \alpha z\frac{\partial}{\partial z} \,.
   \end{align*}
Subsequently we identify this algebra with $\mathcal{K}_{3}=\{X_{1},X_{2},X_{3}\}$, corresponding to the basic isometry algebra of the underlying Bianchi V spacetime.

An action $\mathcal{A}=\int L \,\mbox{d}s$, of the geodesic motion possesses the Lagrangian $L=L(s,x^{\mu},\dot{x}^{\mu})$, in the background of Bianchi V spacetimes, equivalently
 \begin{equation}
 L = \dot{t}^2 - A^2\dot{x}^2 -e^{\alpha x}(B^2\dot{y}^2+C^2\dot{z}^2), \label{lag}
 \end{equation}
 where an over dot represents the derivative with respect to the geodetic parameter $s$. A Noether symmetry is a vector field\footnote{Note that the action of the underlying Lie group is extended to the product manifold $M\times \mathbb{R}$, which is $(4+1)-$dimensional so is to bring the dynamical symmetries (Lie and Noether) and geometrical symmetries (CKV,CC,PC etc) on equal footing. The action naturally includes the parametrization with respect to the geodetic parameter $s$, where this line of approach was followed in \cite{prince}.}
 \begin{align}
 X = \xi \frac{\partial}{\partial s} + \eta^{0} \frac{\partial}{\partial t} + \eta^{1} \frac{\partial}{\partial x}
 + \eta^{2} \frac{\partial}{\partial y}+ \eta^{3} \frac{\partial}{\partial z} \,, \label{sym}
 \end{align}
which leaves the action invariant such that it satisfies the condition
\begin{equation}
X^{(1)} L + L(D_{s} \xi) = D_{s} G\,, \label{cond}
\end{equation}
where $\xi=\xi(s,x^{\mu})$, $\eta^{\mu}=\eta^{\mu}(s,x^{\mu}), ~ (\mu=0,..,3)$ and $G(s,x^{\mu})$ is an unknown function of the corresponding symmetry. The first integrals of the equations of motion are determined by famous Noether theorem which gives an explicit formula \cite{noe}
\begin{equation}
I = \xi	L + (\eta^{\mu} - \dot{x}^{\mu}\xi) \frac{\partial L}{\partial \dot{x}^{\mu}} - G , \label{inv}
\end{equation}
where $x^{\mu}=(t,x,y,z)$ and $\eta^{\mu} = (\eta^{0},\eta^{1},\eta^{2},\eta^{3})$ denote the coefficients of the Noether symmetry (\ref{sym}). Note that $D_{s}I=0$, upon using the equations of motion, where
\begin{equation}
D_{s} = \frac{\partial}{\partial s} + \dot{t}\frac{\partial}{\partial t} +\dot{x}\frac{\partial}{\partial x} +
\dot{y}\frac{\partial}{\partial y}+
\dot{z}\frac{\partial}{\partial z}~.
\end{equation}
The equations of motion (geodesic equations) for the geodesic Lagrangian in Bianchi V spacetimes comprise of a system of four second-order ordinary differential equations in the field variables
  \begin{align}
 &\ddot{t} + e^{\alpha x} ( B {B}^{\prime} ~\dot{y}^2 +C {C}^{\prime} ~\dot{z}^2) + A {A}^{\prime}~\dot{x}^2 = 0, \nonumber\\
 &\ddot{x} + \frac{2A^{\prime}}{A} ~ \dot{x} \dot{t} - \frac{  \alpha e^{\alpha x} (B^2 \dot{y}^2 + C^2 \dot{z}^2 ) }{2A^2}=0, \nonumber\\
 &\ddot{y} + \left (\frac{2B^{\prime}}{B}~ \dot{t} + \alpha \dot{x} \right ) \dot{y}=0, \nonumber\\
 &\ddot{z} + \left (\frac{2C^{\prime}}{C} ~\dot{t} + \alpha \dot{x} \right ) \dot{z}=0, \label{eom}
 \end{align}
where ${}^{\prime}$ represents the derivative with respect to $t$. It is known that Noether algebra is a subalgebra of the Lie algebra of point symmetries of the differential equations \cite{olver}, therefore our study yields classification of Noether algebras of the equations of motion for the geodesic Lagrangian in Bianchi V spacetimes. In order to find Lie point symmetries of the differential equations we extend the vector field (\ref{sym}) to the jet space of second-order where we require that
\begin{equation}
X^{[2]} \bf{E}  = 0 \quad \mbox{mod} \quad \bf{E} \equiv  0,
\end{equation}
where $\bf{E}$, is a given system of differential equations and $X^{[2]}$ is the second-order prolongation which is defined below
\begin{equation}
X^{[2]} = X +  \overset{~[1]~}{~\eta^{\,\mu}} \frac{\partial}{\partial \dot{x}^{\mu}} +  \overset{~[2]~}{~\eta^{\,\mu}} \frac{\partial}{\partial \ddot{x}^{\mu}}  , \label{prolong}
\end{equation}
where $\overset{~[1]~}{~\eta^{\,\mu}}$ and $\overset{~[2]~}{~\eta^{\,\mu}}$ are determined from the formulas
\begin{align}
&\overset{~[1]~}{~\eta^{\,\mu}} = \frac{\mbox{d} \eta^{\mu}}{\mbox{d}s} - \dot{x}^{\mu}\frac{\mbox{d} \xi}{\mbox{d}s}\, , \\
&\overset{~[2]~}{~\eta^{\,\mu}} = \frac{\mbox{d} \overset{[1]}{\eta^{\,\mu}} }{\mbox{d}s} - \ddot{x}^{\mu}\frac{\mbox{d} \xi}{\mbox{d}s} \,.
\end{align}
The Lie symmetries are obtained by applying the operator (\ref{prolong}) on the geodesic equations while replacing the second order derivative terms in the last equation from the equations of motion (\ref{eom}).

Since all Bianchi V spacetimes admit $\mathcal{K}_{3}$ therefore the minimum dimension of the Noether algebra of the geodesic Lagrangian can be obtained easily. Indeed it is easy to prove that the minimal Noether algebra is $\mathcal{N}_{4}=
\mathcal{K}_{3}\oplus \{\partial_{s}\}$, which is of dimension four.

 \textbf{Theorem 1.} \textit{The minimum dimension of Lie algebra of Noether symmetries of the geodesic Lagrangian (\ref{lag}) that leaves the action invariant in Bianchi V spacetimes is 4.}\newline
  \textbf{Proof.} The Lie algebra of isometries, $\mathcal{K}_{3}$, leaves the action invariant therefore all KVs are Noether symmetries of the Lagrangian (\ref{lag}). It is easy to see that the global action of full isometry group $G_{3}$ on the Lagrangian (\ref{lag}) is
 \begin{align*}
 L(s,t,x,y,z) \longrightarrow L(s,t,x+2\epsilon_{3},e^{-\alpha \epsilon_{3}}y+\epsilon_{1},e^{-\alpha \epsilon_{3}}z+\epsilon_{2}),
 \end{align*}
which clearly leaves the action invariant. Therefore all KVs are Noether symmetries of the Lagrangian (\ref{lag}). Finally Lagrangian (\ref{lag}) does not explicitly depend on $s$ therefore invariance under $s-$translation is trivial, i.e., $L(s,t,x,y,z) \longrightarrow L(s+\epsilon_{4},t,x,y,z)$.

 \textbf{Theorem 2.} \textit{The minimum dimension of algebra of Lie point symmetries of the geodesic equations (\ref{eom}) in Bianchi V spacetimes is 5.}\newline
\textbf{Proof.} Since Noether algebra is a subalgebra of Lie symmetry algebra therefore $\mathcal{N}_{4}$ is a Lie subalgebra. Besides the equations of motion (\ref{eom}) are invariant under scaling symmetry $s\partial_{s}$, therefore for all Bianchi V spacetimes the minimum dimension of Lie algebra is five which is identified as $\mathcal{N}_{4}\oplus \{s\partial_{s}\}$.

In order to identify Bianchi V spacetimes that possess minimal Noether algebra $\mathcal{N}_{4}$ of dimension four and other algebras we solve the determining equations for Noether symmetries which are obtained by splitting equation (\ref{cond}) in monomials and obtain a set of nineteen linear partial differential equations
 \begin{align*}
 &\xi_{t}=0,~ \xi_{x}=0, ~ \xi_{y} =0, ~\xi_{z}=0,~ \xi_{s} -2 \eta_{0t} =0,
\\
&\eta_{0x} -A^2 \eta_{1t} =0,~ \eta_{0y} -e^{\alpha x} B^2 \eta_{2t} =0, ~  \eta_{0z} - C^2 e^{\alpha x} \eta_{3t} =0,  \\
&A^2 \eta_{1y} +B^2 e^{\alpha x} \eta_{2x}=0, ~ B^2\eta_{2z} +C^2 \eta_{3y}=0, A^2 \eta_{1z} + C^2 e^{\alpha x} \eta_{3x}=0 \\
&  2 \eta_{0} A^{\prime}+ A(2 \eta_{1x}-\xi_{s}) =0, \\
& 2\eta_{0}B^{\prime} + B( \alpha \eta_{1} -\xi_{s} +2\eta_{2y})=0,
\\
&2\eta_{0} C^{\prime} + C( \alpha \eta_{1} -\xi_{s} +2\eta_{3z})=0, ~  \\
&G_{s} =0 , ~ G_{t}-2\eta_{0s} =0,
\\
&G_{x} + 2 A^2 \eta_{1s} =0, ~ G_{y}  + 2 B^2 e^{\alpha x} \eta_{2s} =0, ~ G_{z} + 2C^2 e^{\alpha x} \eta_{3s} =0.
 \end{align*}
The above system can be integrated for various forms of the arbitrary functions. We use CAS (Maple) to categorize all the cases using the command `rifsimp'. It is interesting to see that in most cases the function $A(t)$ turns out to be an affine function. Therefore the above system of PDEs uniquely characterize the affine form of $A(t)$ which specify the forms of other arbitrary functions. We denote Noether algebra with $\mathcal{N}_{i}$, where $i$ refers to its dimensions and the distinction of subcases is made through another subscript. We observe that the following cases arise and in each case we also obtain the first integrals of the equations of motion (\ref{eom}) using Noether theorem (\ref{inv}).

\textbf{Case 1.} (4-dimensional algebra)
 \newline
 It is identified as $\mathcal{N}_{4}$, which arise for the geodesic Lagrangian in Bianchi V spacetimes where evolutionary functions are arbitrary and not among those that arise in the subsequent cases.
%  \begin{align*}
%  & \textbf{1a.} \quad A(t)= c_{1}t +c_{2}, \quad B(t) = c_{3}  ,\quad C(t) \neq A(t) \neq c_{4}, \\
%  & \textbf{1b.} \quad  A(t)= c_{1}t +c_{2}  , \quad C(t) = c_{3}, \quad B(t) \neq A(t)\neq c_{4}  , \\
%  & \textbf{1c.} \quad  B(t)=c_{1}t+c_{2}, \quad A(t) \neq B(t) \neq C(t)  ,	\\
%  & \textbf{1d.} \quad C(t)=c_{1}t+c_{2}, \quad A(t) \neq C(t)  \neq B(t) ,
%  \end{align*}
The four independent invariants of the equations of motion corresponding to $\mathcal{N}_{4}$, include three invariants for algebra $\mathcal{K}_{3}$ and fourth invariant is for $\partial_{s}$, which is the Lagrangian itself, or equivalently
  \begin{align}
I_{1} = B^2e^{\alpha x} \dot{y},\quad I_{2} = C^2e^{\alpha x}\dot{z},\quad I_{3} = 2A^2\dot{x} -\alpha e^{\alpha x}(B^2y\dot{y}+ C^2z \dot{z}) ,\quad I_{4} = L. \label{inv1}
\end{align}
The Lie algebra of point symmetries of the geodesic equations includes all Noether symmetries and an additional scaling symmetry $\mathcal{L}_{1}=s\partial_{s}$, which is not a Noether symmetry. Here we use $\mathcal{L}$, to denote a Lie symmetry which is not a Noether symmetry with respect to a geodesic Lagrangian. Therefore, the Lie algebra is $\mathcal{N}_{4}\oplus \{\mathcal{L}_{1}\}$ and is of dimensions five.

 \textbf{Case 2.} (5-dimensional algebra)
 \newline
 There are three cases of five dimensional Noether algebras which are possessed by the Lagrangian in Bianchi V spacetimes with
 \begin{align*}
 &\textbf{2a.} \quad  A(t) = c_{1}t, \quad B(t)=c_{2}t^{m}, \quad C(t)= c_{3}t^{n}, \quad m\neq n \,,\\
 &\textbf{2b.} \quad  C(t)=c_{1}B(t) , \quad B(t)\neq A(t)\,, \quad C(t) \neq \mbox{const.} \\
 &\textbf{2c.} \quad  A(t) = c_{1}, \quad B(t)=c_{2}e^{\beta t}, \quad C(t)= c_{3}e^{\gamma t}, \quad \beta \neq \gamma  \,,
  \end{align*}
where $c_{i}, ~ \forall ~i$ are non-zero constants. In the first case both powers $m$ and $n$ can not be equal to zero or  one, simultaneously. Similarly the powers $\beta$ and $\gamma$ can not be zero, simultaneously in case (2c). In the first case we obtain two Noether algebras
\begin{align}
&\mathcal{N}_{5,a}=\mathcal{N}_{4}\oplus \{2\alpha s\partial_{s}+ \alpha t\partial_{t}+2(1-m)\partial_{x}+\alpha (m-n)z\partial_{z}\}, \\
&\mathcal{N}_{5,a}=\mathcal{N}_{4}\oplus \{2\alpha s\partial_{s}+ \alpha t\partial_{t}+2(1-n)\partial_{x}-\alpha (m-n)y\partial_{y}\}, \\
&\mathcal{N}_{5,b}=\mathcal{N}_{4}\oplus \{c_{1}^2z\partial_{y}-y\partial_{z} \} \,,\\
&\mathcal{N}_{5,c}=\mathcal{N}_{4}\oplus \{ \partial_{t}-\beta y\partial_{y}-\gamma z \partial_{z} \} \,,
\end{align}
Besides invariants $I_{1} - I_{4}$, an additional invariant for each case is
\begin{align}
&I_{5,a} = -e^{\alpha x} \left (((m-n)z-s\dot{z} )c_{3}^2\dot{z}t^{2n}-c_{2}^2s\dot{y}^2 t^{2m} \right ) +\frac{2(m-1)}{\alpha}c_{1}^2t^2 \dot{x} + t\dot{t}-1+ \nonumber
 \\
&\quad\quad  s\left (c_{1}^2t^2\dot{x}^2-\dot{t}^2 \right) \,, \\
&I_{5,a} = e^{\alpha x} \left (((m-n)y+s\dot{y} )c_{2}^2\dot{y}t^{2m}+c_{3}^2s\dot{z}^2 t^{2n} \right )+\frac{2(n-1)}{\alpha}c_{1}^2t^2 \dot{x} + t\dot{t}-1+\nonumber
 \\
&\quad  \quad s\left (c_{1}^2t^2\dot{x}^2+\dot{t}^2 \right) \,,
\end{align}
however for other two cases the invariants include
\begin{align}
&I_{5,b} = B^2c_{1}^2e^{\alpha x} (y\dot{z}-z\dot{y}) \,, \label{5c5}\\
&I_{5,c} =  \dot{t} + c_{2}^2 \beta y \dot{y} e^{\alpha x +2 \beta t} + c_{3}^2 \gamma z \dot{z} e^{\alpha x + 2\gamma t}\, .
\end{align}
The algebra of Lie point symmetries is six in all cases where the identification is $\mathcal{N}_{5,I}\oplus \{\mathcal{L}_{1}=s\partial_{s} \},$ where $I\in \{a,b,c\}$.

 \textbf{Case 3.} (6-dimensional algebra)
 \newline
We obtain two cases in which the geodesic Lagrangian admits a $6-$dimensional Noether algebra. These are identified as
\begin{align}
&\mathcal{N}_{6,a}=\mathcal{N}_{5,b}\oplus \{2\alpha s\partial_{s}+\alpha t\partial_{t}+2(1-n)\partial_{x}\} \\,
&\mathcal{N}_{6,b}=\mathcal{N}_{5,b}\oplus \{2c_{1}\alpha s\partial_{s}+\alpha (c_{1}t+c_{2})\partial_{t}+2c_{1}\partial_{x}\}\,,
\end{align}
up to a redefinition of constant in case 2, which are possessed by the Lagrangians in Bianchi V spacetimes
 \begin{align}
 &\textbf{3a.} \quad A(t)= c_{1}t, \quad B(t) = c_{2}t^{n}, \quad C(t)=c_{3}t^{n}, \quad n\neq 1 \,.\\
 &\textbf{3b.} \quad A(t)= c_{1}t+c_{2}, \quad B(t) = c_{3}, \quad C(t)=c_{4}\,,
 \end{align}
respectively. The invariants $I_{1}-I_{5}$ are same as before with an extra invariant
\begin{align}
&I_{6,a} = \frac{2(n-1)c_{1}^2 t^2 \dot{x}}{\alpha} + \left ( t\dot{t} -1 + s (c_{1}^2t^2 \dot{x}^2-\dot{t}^2 ) \right )+ se^{\alpha x} t^{2n}(c_{2}^2 \dot{y}^2 + c_{3}^2 \dot{z}^2)\,, \\
&I_{6,b}= \alpha c_{2} \dot{t} + c_{1} \left (\alpha (s(c_{2}^2 \dot{x}^2-\dot{t}^2)+t\dot{t}-1) - 2c_{2}^2 \dot{x} \right ) + c_{1}^2 t(2c_{2} \dot{x}+c_{1} t\dot{x})(s\alpha \dot{x}-2)+ \\
&\quad \quad c_{1}\alpha s  (c_{3}\dot{y}^2 +c_{4}^2 \dot{z}^2)e^{\alpha x}\,.
\end{align}
We find that there are seven Lie symmetries for the geodesic equations where we get a decomposition of the extra vector fields obtained in each case. In particular, the vector field $2\alpha s\partial_{s}+\alpha t\partial_{t}+2(1-n)\partial_{x}$, divides into two independent Lie symmetries
\begin{equation}
\mathcal{L}_{1,a} = s \frac{\partial}{\partial s}, \quad \mathcal{L}_{2,a} = \alpha t \frac{\partial}{\partial t} + 2(1-n) \frac{\partial}{\partial x} \,,
\end{equation}
such that there linear combination is a Noether symmetry but independently these are not Noether symmetries. It is easy to check that the algebra $\mathcal{N}_{5,b} \oplus \{\mathcal{L}_{1,a},\mathcal{L}_{2,a}\}$, is a seven-dimensional Lie algebra of point symmetries for equations of motion in case (3a). For the other case (3b)
the Lie algebra of point symmetries is $\mathcal{N}_{5,b} \oplus \{\mathcal{L}_{1,b},\mathcal{L}_{2,b}\}$, where
\begin{equation}
\mathcal{L}_{1,b} = s \frac{\partial}{\partial s}, \quad \mathcal{L}_{2,b} = (c_{1}t+c_{2})  \frac{\partial}{\partial t} + \frac{2c_{1}}{\alpha} \frac{\partial}{\partial x} \,.
\end{equation}

 \textbf{Case 4.} (7-dimensional algebra)
 \newline
 The geodesic Lagrangian with all evolutionary functions being equal
 \begin{align}
 A(t)= B(t)=C(t) \neq c_{1}t+c_{2} \, ,
 \end{align}
 admits $\mathcal{N}_{5,b}$ algebra along with two additional symmetries
  \begin{align*}
 &
X_{6} = z\frac{\partial}{\partial x} -\frac{\alpha yz}{2} \frac{\partial}{\partial y} + \frac{1}{4 \alpha}
\left (4 e^{-\alpha x} + \alpha^2 (y^{2}-z^2) \right)\frac{\partial}{\partial z} \, ,\\
&X_{7} = y\frac{\partial}{\partial x}+ \frac{1}{4 \alpha}\left (4e^{-\alpha x} - \alpha^2 (y^{2}-z^2) \right)\frac{\partial}{\partial y}  -\frac{\alpha yz}{2} \frac{\partial}{\partial z}\, ,
 \end{align*}
therefore $\mathcal{N}_{7}=\mathcal{N}_{5,b}\oplus \{X_{6},X_{7}\}$. The two new invariants are
\begin{align}
&I_{6} = A^2 \left ( 4\dot{z}  + 4\alpha z \dot{x} + \alpha^2 e^{\alpha x}\left(  (y^2 - z^2) \dot{z} -2yz \dot{y} \right)\right), \\
&I_{7} = A^2 \left ( 4\dot{y}  + 4\alpha y \dot{x} - \alpha^2 e^{\alpha x}\left(  (y^2 - z^2) \dot{z} -2yz \dot{z} \right)\right).
\end{align}
The Lie algebra of point symmetries is $\mathcal{N}_{7}\oplus \{\mathcal{L}_{1}=s\partial_{s}\},$ which has dimension eight.

 \textbf{Case 5.} (9-dimensional algebra)\newline
In this case there appear three subcases in which all evolutionary functions are specified. In the first two subcases, $\mathcal{N}_{7}$ is a subalgebra of $\mathcal{N}_{9}$, however in the last subcase only $\mathcal{N}_{5,b}$ is a subalgebra.

 \textbf{5a.} It is identified as $\mathcal{N}_{9,a}=\mathcal{N}_{7} \oplus \{X_{8},X_{9}\}$, where evolutionary functions are specified by
 \begin{align}
 A(t)=c_{1}, \quad B(t) = c_{2},\quad C(t)=c_{3}.
  \end{align}
 Additional symmetries include
 \begin{align*}
&
X_{8} = \frac{\partial}{\partial t} , \quad  X_{9}= s\frac{\partial}{\partial t} \,,
 \end{align*}
 where the gauge term is constant in all cases except for $X_{9}$, which is obtained with non-constant gauge $G=2t$. The corresponding conserved quantities are
 \begin{align}
I_{8} = \dot{t} , \quad I_{9}= s\dot{t} - t  \,.
\end{align}
The Lie symmetry algebra of the corresponding geodesic equations is of twelve dimension which is identified as $\mathcal{N}_{7} \oplus \{X_{8}, X_{9}\} \oplus \{\mathcal{A}_{3}^{\mathcal{L}}\}$, where $\mathcal{A}_{3}^{\mathcal{L}}$ is
\begin{equation}
\mathcal{L}_{1}=s\frac{\partial}{\partial s}, \quad \mathcal{L}_{2}=t\frac{\partial}{\partial t},\quad \mathcal{L}_{3}=t\frac{\partial}{\partial s} \,.
\end{equation}

  \textbf{5b.} The subalgebra is $\mathcal{N}_{9,b}=\mathcal{N}_{7} \oplus \{X_{8},X_{9}\}$, for
 \begin{align}
 A(t)=c_{1}t+c_{2}, \quad B(t) = c_{2}A(t),\quad C(t)= c_{3}A(t) \,.
  \end{align}
 Additional symmetries include
 \begin{align*}
 &X_{8}= s^2\frac{\partial}{\partial s}+\frac{s(c_{1}t+c_{2})}{c_{1}}\frac{\partial}{\partial t}\,, \quad G=\frac{c_{1}t^2+2c_{2}t}{2c_{1}} \,,\\
 &
X_{9}= 2s\frac{\partial}{\partial s}+\frac{c_{1}t+c_{2}}{c_{1}}\frac{\partial}{\partial t}\,, \quad G=C_{1},
  \end{align*}
  with invariants
  \begin{align}
&I_{8} = c_{1}s^2 (c_{1}t+c_{2})^2 (c_{3}^2\dot{y}^2 +c_{4}^2 \dot{z}^2) e^{\alpha x} + c_{1}^2 t s^2 \dot{x}^2
(c_{1} t +2 c_{2}) +2c_{2} (s\dot{t}-t)+ \\
& \quad \quad  +c_{1} \left( (c_{2}^2\dot{x}^2-\dot{t}^2)s^2 + 2st\dot{t}-t^2\right) \,, \\
&I_{9}=  c_{1}s (c_{1}t+c_{2})^2 (c_{3}^2\dot{y}^2 + c_{4}^2\dot{z}^2) e^{\alpha x} +c_{1}^2 t s \dot{x}^2
(c_{1} t +2 c_{2}) + c_{1} ( (c_{2}^2\dot{x}^2-\dot{t}^2)s + t\dot{t})+c_{2}\dot{t}  \,.
\end{align}
Now in this case the algebra of Lie point symmetries is only ten dimensional unlike the previous case $(5a)$, where the extra symmetry generator arise from the decomposition of $X_{9}$, i.e.
\begin{equation}
\mathcal{L}_{1} = s \frac{\partial}{\partial s}, \quad \mathcal{L}_{2} = (c_{1}t+c_{2}) \frac{\partial}{\partial t} \,.
\end{equation}
The Lie algebra $\mathcal{N}_{7}\oplus  \{X_{8}\} \oplus \{\mathcal{L}_{1} ,\mathcal{L}_{2} \}$, is closed which is of dimension ten.

\textbf{5c.} It is identified as $\mathcal{N}_{9,c}=\mathcal{N}_{5,b}\oplus \mathcal{A}^{\mathcal{N}}_{4}$, for
 \begin{align}
 A(t)=c_{1}, \quad B(t) = c_{2}e^{\beta t},\quad C(t)=c_{3}e^{\beta t}, \quad \beta \neq \pm\,\frac{\alpha }{2c_{1}}
  \end{align}
 where $\mathcal{A}^{\mathcal{N}}_{4}$ refers to four additional Noether symmetries
 \begin{align*}
&
X_{6}=\frac{\partial}{\partial t} -\beta y\frac{\partial}{\partial y} -\beta z \frac{\partial}{\partial z} \, , \\
&
X_{7}= \frac{\alpha s}{2\beta} \frac{\partial}{\partial t}-s\frac{\partial}{\partial x}, \quad G=x+ \frac{\alpha t}{2\beta c_{1}^2 }\, ,\\
&X_{8}= -\frac{2c_{1}^2 \beta y}{\alpha} \frac{\partial}{\partial t} + y\frac{\partial}{\partial x} +
\frac{1}{4\alpha c_{2}^2} ((\alpha^2 - 4c_{1}^2\beta^2)  (c_{3}^2 z^2 - c_{2}^2 y^2) +4c_{1}^2 e^{-\alpha x - 2\beta t})\frac{\partial}{\partial y} - \\
& \quad \quad -\frac{yz( \alpha ^2 -4c_{1}^2 \beta^2) }{2\alpha}\frac{\partial}{\partial z} \, , \\
&X_{9}= -\frac{2c_{1}^2 \beta z}{\alpha} \frac{\partial}{\partial t} + z\frac{\partial}{\partial x} + \frac{1}{4\alpha c_{2}^2} ((\alpha^2 - 4c_{1}^2\beta^2)  (c_{3}^2 z^2 - c_{2}^2 y^2) +4c_{1}^2 e^{-\alpha x - 2\beta t})\frac{\partial}{\partial z} - \\
&\quad \quad
-\frac{yz( \alpha ^2 -4c_{1}^2 \beta^2) }{2\alpha}\frac{\partial}{\partial y}  \,.
  \end{align*}
  The corresponding invariants of the geodesic motion are
   \begin{align}
&I_{6} = \dot{t} + \beta e^{\alpha x + 2 \beta t} (c_{2}^2 y\dot{y}+ c_{3}^2 z \dot{z})  , \label{5c1}\\
&I_{7}= 2\beta c_{1}^2 (s \dot{x} - x) + \alpha (s \dot{t} - t)   ,\label{5c2} \\
&I_{8}= (4c_{1}^2 \beta^2 - \alpha ^2) ( c_{2}^2 y^2\dot{y} + c_{3}^2 z(2y\dot{z} - z\dot{y}))e^{\alpha x + 2\beta t}+
4c_{1}^2( y ( \alpha \dot{x} + 2\beta \dot{t} ) + \dot{y}), \label{5c3}\\
&I_{9}= (4c_{1}^2 \beta^2 - \alpha ^2) ( c_{3}^2 z^2\dot{z} + c_{2}^2 y (2z\dot{y} - y\dot{z}))e^{\alpha x + 2\beta t}+
4c_{1}^2( y ( \alpha \dot{x} + 2\beta \dot{t} ) + \dot{z}) \label{5c4}.
\end{align}
In this case we again obtain twelve dimensional Lie algebra of the geodesic equations which is $\mathcal{N}_{5,b}\oplus \{X_{8},X_{9}\}\oplus \mathcal{A}_{5}^{\mathcal{L}}$, where $\mathcal{A}_{5}^{\mathcal{L}}$ is
\begin{align*}
&{\mathcal{L}}_{1} = s \frac{\partial}{\partial {s}}, \quad {\mathcal{L}}_{2} = \left(x + \frac{\alpha t}{2c_{1}^2 \beta}\right) \frac{\partial}{\partial {s}}, \quad {\mathcal{L}}_{3} = \frac{\partial}{\partial {t}} - \frac{2\beta }{\alpha } \frac{\partial}{\partial {x}}, \\
&{\mathcal{L}}_{4} = s \mathcal{L}_{3}, \quad {\mathcal{L}}_{5} = \left(x + \frac{\alpha t}{2c_{1}^2 \beta}\right) \frac{\partial}{\partial {t}} - \frac{2\beta}{\alpha}\left(x + \frac{\alpha t}{2c_{1}^2 \beta}\right) \frac{\partial}{\partial {x}}\,.
\end{align*}

\textbf{Case 6.} (10-dimensional algebra)\newline
There are two subcases of ten dimensional algebras of Noether symmetries corresponding to $\alpha =  2c_{1}\beta $ and $\alpha = - 2c_{1}\beta $, respectively.

\textbf{6a.} It is identified as $\mathcal{N}_{9,c}\oplus {\mathcal{A}_{1}^\mathcal{N}}$, with
 \begin{align}
 A(t)=c_{1}, \quad B(t) = c_{2}e^{\beta t},\quad C(t)=c_{3}e^{\beta t}, \quad \beta= \frac{\alpha}{2c_{1}}
  \end{align}
 where ${\mathcal{N}_{9}}$ is the same as in case (5c) by substituting the value $\alpha =  2c_{1}\beta$. An additional symmetry ${\mathcal{A}_{1}^\mathcal{N}}$ that arises is
 \begin{align*}
 &
X_{10}=2s\frac{\partial}{\partial s}+ (t - c_{1}x)\frac{\partial}{\partial t} -  \frac{t-c_{1}x}{c_{1}} \frac{\partial}{\partial x}+ y\frac{\partial}{\partial y}+ z\frac{\partial}{\partial z} \, ,
  \end{align*}
 with corresponding invariant
   \begin{align}
I_{10} = (c_{1}\dot{x} + \dot{t}) ( c_{1}(s\dot{x} -x) + ( t- s\dot{t}) )+ c_{1}^2 \left (s(c_{2}^2\dot{y}^2 + c_{3}^2\dot{z}^2) -
c_{2}^2y \dot{y} -c_{3}^2 z \dot{z} \right )e^{2\beta (t + c_{1}x )}.
\end{align}
In this case we obtain thirteen dimensional Lie algebra of the geodesic equations which is $\mathcal{N}_{5,b}\oplus \{X_{8},X_{9}\}\oplus \mathcal{A}_{6,a}^{\mathcal{L}}$, where $\mathcal{A}_{6,a}^{\mathcal{L}}$ is
\begin{align*}
&{\mathcal{L}}_{1} = s \frac{\partial}{\partial {s}}, \quad {\mathcal{L}}_{2} = \left(x + \frac{t}{c_{1}}\right) \frac{\partial}{\partial {s}}, \quad {\mathcal{L}}_{3} = \frac{\partial}{\partial {t}} - \frac{1}{ c_{1}} \frac{\partial}{\partial {x}}, \\
&{\mathcal{L}}_{4} = s{\mathcal{L}}_{8}, \quad {\mathcal{L}}_{5} = t\frac{\partial}{\partial {t}} + \frac{1-2\beta t}{2c_{1} \beta}\frac{\partial}{\partial {x}}, \quad {\mathcal{L}}_{6} = x\frac{\partial}{\partial {t}} - \frac{1+2c_{1}\beta t}{2c_{1}^2 \beta}\frac{\partial}{\partial {x}}\,.
\end{align*}

\textbf{6b.} It is identified as $\mathcal{N}_{9,c}\oplus {\mathcal{A}_{1}^\mathcal{N}}$, with
 \begin{align}
 A(t)=c_{1}, \quad B(t) = c_{2}e^{\beta t},\quad C(t)=c_{3}e^{\beta t}, \quad \beta = -\frac{\alpha}{2c_{1}}
  \end{align}
 where ${\mathcal{N}_{9}}$ is the same as in case (5c) by substituting the value $\alpha = - 2c_{1}\beta$. An additional symmetry ${\mathcal{A}_{1}^\mathcal{N}}$ that arise is
 \begin{align*}
 &
X_{10}=2s\frac{\partial}{\partial s}+ (t+ c_{1}x)\frac{\partial}{\partial t} + \frac{t-c_{1}x}{c_{1}} \frac{\partial}{\partial x}+ y\frac{\partial}{\partial y}+ z\frac{\partial}{\partial z} \, ,
  \end{align*}
 with corresponding invariant
   \begin{align}
I_{10} = (c_{1}\dot{x} - \dot{t}) ( c_{1}(s\dot{x} -x)-( t- s\dot{t}) )+ c_{1}^2 \left (s(\dot{y}^2 + \dot{z}^2) -
y \dot{y} -z \dot{z} \right )e^{2\beta (t-c_{1}x )}.
\end{align}
In this case we again obtain thirteen dimensional Lie algebra of the geodesic equations which is $\mathcal{N}_{5,b}\oplus \{X_{8},X_{9}\}\oplus \mathcal{A}_{6,b}^{\mathcal{L}}$, where $\mathcal{A}_{6,b}^{\mathcal{L}}$ is
\begin{align*}
&{\mathcal{L}}_{1} = s \frac{\partial}{\partial {s}}, \quad {\mathcal{L}}_{2} = \left(x - \frac{t}{c_{1}}\right) \frac{\partial}{\partial {s}}, \quad {\mathcal{L}}_{3} = \frac{\partial}{\partial {t}} - \frac{1}{ c_{1}} \frac{\partial}{\partial {x}}, \\
&{\mathcal{L}}_{4} = s{\mathcal{L}}_{8}, \quad {\mathcal{L}}_{5} = t\frac{\partial}{\partial {t}} - \frac{1-2\beta t}{2c_{1} \beta}\frac{\partial}{\partial {x}}, \quad {\mathcal{L}}_{6} = x\frac{\partial}{\partial {t}} - \frac{1-2c_{1}\beta t}{2c_{1}^2 \beta}\frac{\partial}{\partial {x}}\,.
\end{align*}

This completes the classification of Noether symmetries of the geodesic Lagrangian in Bianchi V spacetimes. The equations of motion (\ref{eom}) inherit a twelve dimensional Lie algebra of Lie point symmetries for the simplest case, in which all three scale factors are constant which is not the maximal algebra as one would expect for the simplest model. The reason why simplest Bianchi V model does not attain a maximal algebra of dimension $13$, does not come as a surprise because the underlying Riemannian manifold is not flat but contains flat sections. The algebra of Lie symmetries of the equations of motion in a flat space is unique and corresponds to $sl(n+2,R)$ \cite{tsam}. In our analysis, Bianchi V spacetimes are not flat in general (except when $\alpha =0$) the maximum dimension of Lie algebra of Lie point symmetries is thirteen. Noether symmetry classification of the geodesic Lagrangian in Bianchi V spacetimes reveal that there are six Noether algebras of dimensions $4,5,6,7,9$ or $10$ which are the subalgebras of the $13-$dimensional Lie algebra. Thus, we have established the following results.

\textbf{Proposition 1.} \quad \textit{The geodesic Lagrangian (\ref{lag}) of Bianchi V spacetimes can have Noether algebra of dimensions $4,5,6,7$, $9$ or $10$. }

\textbf{Proposition 2.} \quad \textit{The algebra of Lie point symmetries of the geodesic equations of Bianchi V spacetimes specified by the Noether symmetries, can have dimensions $5,6,7,8$, $10$, $12$ or $13$. }

The summary of our results is given in Table 1, where $d(\mathcal{N})$ and $d(\mathcal{L})$ refers to the dimensions of Noether and Lie algebras, respectively.
\begin{table}
\caption {Dimensions of Noether and Lie algebras of point symmetries}
\begin{tabular}{| l |c| c| c| r |}
\hline
 \textbf{Cases} & \textbf{Noether Algebra} $\mathcal{N}$ & $d(\mathcal{N})$ & \textbf{Lie Algebra} $\mathcal{L}$ & $d(\mathcal{L})$\\ \hline 	\hline
  1 & $\mathcal{N}_{4}$ & 4 & $\mathcal{N}_{4}\oplus \{\mathcal{L}_{1}\}$  &5 \\\hline
  2a & $\mathcal{N}_{5,a}$  &5 & $\mathcal{N}_{5,a}\oplus \{\mathcal{L}_{1}\}$& 6
  \\\hline
  2b & $\mathcal{N}_{5,b}$ &5 & $\mathcal{N}_{5,b}\oplus \{\mathcal{L}_{1}\}$& 6
  \\\hline
  2c & $\mathcal{N}_{5,c}$ &5 & $\mathcal{N}_{5,c}\oplus \{\mathcal{L}_{1}\}$& 6
 \\\hline
  3a & $\mathcal{N}_{6,a}$ & 6  & $\mathcal{N}_{5,b} \oplus \{\mathcal{L}_{1,a},\mathcal{L}_{2,a}\}$&7 \\\hline
  3b & $\mathcal{N}_{6,b}$& 6  & $\mathcal{N}_{5,b} \oplus \{\mathcal{L}_{1,b},\mathcal{L}_{2,b}\}$&7 \\\hline
  4 &$\mathcal{N}_{7}$& 7 & $\mathcal{N}_{7}\oplus \{\mathcal{L}_{1}\}$ &8 \\\hline
  5a &$\mathcal{N}_{9,a}$ & 9 & $\mathcal{N}_{7} \oplus \{X_{8}, X_{9}\} \oplus \mathcal{A}_{3}^{\mathcal{L}}$& 12  \\\hline
  5b & $\mathcal{N}_{9,b}$& 9 &$\mathcal{N}_{7}\oplus  \{X_{8}\} \oplus \{\mathcal{L}_{1} ,\mathcal{L}_{2} \}$& 10 \\\hline
   5c & $\mathcal{N}_{9,c}$& 9 &$\mathcal{N}_{5,b}\oplus \{X_{8},X_{9}\}\oplus \mathcal{A}_{5}^{\mathcal{L}}$ & 12 \\\hline
  6a & $\mathcal{N}_{10,a}$ & 10 & $\mathcal{N}_{5,b}\oplus \{X_{8},X_{9}\}\oplus \mathcal{A}_{6,a}^{\mathcal{L}}$& 13 \\
 \hline
  6b & $\mathcal{N}_{10,b}$ & 10 & $\mathcal{N}_{5,b}\oplus \{X_{8},X_{9}\}\oplus \mathcal{A}_{6,b}^{\mathcal{L}}$& 13 \\
\hline
\end{tabular}
\end{table}

 \section{Physical Interpretation of New Solutions}
The study of inhomogeneous and anisotropic cosmologies started soon after the birth of GR \cite{kasner} and a detailed analysis of exact solutions in terms of the asymptotic of singularities appeared in the works \cite{lif1,lif2} followed by numerous attempts \cite{bergh,wain,tolman,mat}. For a more detailed survey on such cosmologies the reader is referred to \cite{mac}. The class of Bianchi V spacetimes contains anisotropic and homogeneous cosmologies which are crucial to investigate for several reasons. The universe is homogeneous at very large scales and the question of whether it had started with a little bit irregularity require us to analyze small perturbations away from the high symmetry of Friedmann models. This could help us investigate on the present-day anisotropy of the microwave background radiation and irregularities in density and temperature at the early epochs when the radiation was emitted. Besides adiabatic cooling, viscous dissipation and particle creation of an anisotropic universe can be studied with new cosmological models.

The information of matter content in Bianchi V spacetimes is contained in the Riemann and Weyl tensors which can be used to examine the effect of tidal forces due to curvature in the manifold along the geodesics. The former helps us to monitor the change in volume while later provides change in shape of the observer along geodesics. The basic requirement for a spacetime to be physical is that the positive energy condition is met. The positive energy condition requires that $T_{00}$ which corresponds to the energy density is non-negative therefore $T_{00}\geq 0$. 
It is worth pointing out that the existence of such Noether symmetries completely specify the cosmological models which is in agreement with the results in \cite{cap2}, where point-like Noether symmetries were employed to determine feasible models in extended gravity quantum cosmology. Our prime interest here is to investigate the positive energy condition in the models specified by Noether symmetries. 

The components of Einstein tensor, Weyl and curvature tensors involving arbitrary functions are already given in \cite{cam}. It was also shown that the rank of a $6\times 6$ curvature matrix is $3$, 4, 5 or 6 in Bianchi V spacetimes where the case of rank 3 give rise to infinite dimensional Lie algebra of proper CCs. On the other hand the rank of $6\times 6$ Weyl matrix is 0, 4 or 6 and such spacetimes do not admit proper WCs except for the trivial rank zero case. The Lie algebra of proper conformal KVs in such spacetimes is four-dimensional. We now discuss the physical interpretation of the cosmological solutions obtained in the last section in the light of positive energy condition.

\textbf{Case 1.} ($4-$dimensional algebra)\newline
The class of Bianchi V spacetimes admitting 4-dimensional algebra of Noether symmetries contains all three arbitrary scale factors and the corresponding invariants are given in (\ref{inv1}). We investigate general physical characteristics of this spacetime by considering the energy density
\begin{equation}
\rho(t) = \frac{A^{\prime}}{A} \,  \frac{B^{\prime}}{B}  + \frac{B^{\prime}}{B}\,  \frac{C^{\,\prime}}{C}+  \frac{A^{\prime}}{A}\, \frac{C^{\,\prime}}{C} - \frac{3\,\alpha^2}{4A^2}\,,
\end{equation}
where the positive energy condition requires that $\rho(t)\geq 0$. It is convenient to bring $\rho(t)$ into a more useful form
\begin{equation}
\rho(t) = a^{\prime} b^{\prime} +  b^{\prime} c^{\prime} + a^{\prime} c^{\prime}  - \frac{3\alpha^2}{4} e^{-2a}\,, \label{den}
\end{equation}
by introducing a simple change of quantities $a = \ln A, ~b= \ln B, ~ c = \ln C$. The first three terms in the above equation does have the qualitative behavior of kinetic energies in Newtonian mechanics \cite{misner}. Since $a^{\prime}$, encodes the information how fast or slow the expansion takes place, therefore the mixed term $a^{\prime}b^{\,\prime}$, can be regarded as the kinetic energy of the composite system of both $a$ and $b$. Similar argument holds for the composite systems ($b$,$c$) and ($a$,$c$). However it is important to note that in the above relation we do not have the contribution of individual kinetic energies due to $a$, $b$ and $c$, respectively. In fact the product terms can be set equivalently to
\begin{equation}
\rho(t) = \frac{1}{2}\left ( a^{\prime} + b^{\prime} +   c^{\prime} \right )^2 -\frac{1}{2} \left( a^{\prime 2} +b^{\prime 2} + c^{\prime 2} \right)  - \frac{3\alpha^2}{4} e^{-2a}\,,
\end{equation}
where the first term is positive and corresponds to the kinetic energy $T_{c}$, of the composite system $a$, $b$ and $c$. The second term is the total kinetic energy $T_{e},$ of individual systems which is negative. Therefore in all Bianchi V spacetimes the total kinetic energy of individual and composite systems is irrelevant and the quantity which is crucial is the difference $T_{c}-T_{e}$. For a realistic model we require that $T_{c}>T_{e}$. The energy density is negative in the case when $T_{c}=T_{e}$, where the spacetime is unrealistic. The last term could be regarded as the potential term which is an exponential function of the scale factor $a$. The significance of the first scale factor $A(t)$, upon the others is apparent and notably Noether symmetries also characterized it.

Note that the contribution of the last term is small compared to other terms as long as $|A(t)|$ grows with time. However the evolution of $A(t)$ is critical in the interval $A(t)\in \[0,1\],$ in which case $a(t)$ is negative and the last term plays a significant contribution in decreasing the overall energy density of the spacetime. On the other hand the slopes of $b(t)$ and $c(t)$ play a significant role on the energy density of spacetimes. In particular from the equation (\ref{den}), it is clear that the contribution of the product $b^{\prime}c^{\prime},$ is larger than the sum $b^{\prime}+c^{\prime}$. Therefore if both $b$ and $c$, simultaneously accelerate or decelerate then both slopes are positive or negative and we expect the spacetime to be realistic. Moreover if the slopes are opposite then the only possibility for a physical spacetime is that when
$A(t)\in \[0,1\],$ in which case the last term contribute in a positive energy density so is to balance the effect of other terms. Now we consider those cases in which the evolutionary functions are specified by the existence of Noether symmetries and identify the critical bounds on the expansion parameters for realistic Bianchi V spacetimes.

 \textbf{Case 2.} (5-dimensional algebra)\newline
In this case we obtained three subcases in which one of the subcase contains arbitrary functions. We consider the cases in which all functions are completely specified
\begin{align}
  \textbf{2a.} \quad \mbox{d}s^2 = \mbox{d}t^2 - c_{1}^2 t^2 \mbox{d}x^2 -  e^{\alpha x}  ( c_{2}^2 t^{2m}\mbox{d}y^2 + c_{3}^2 t^{2n}\mbox{d}z^2), \quad m\neq n\,,
  \end{align}
which has non-zero components of the Weyl tensor therefore it is not conformally flat spacetime. There are eleven non-zero curvature invariants where we mention only Ricci scalar and denote it by $I^{\mathcal{R}}$, given by
\begin{equation}
I^{\mathcal{R}}_{1,a}=\frac{3\alpha^2 -4c_{1}^2 (m^2 +mn +n^2)}{2c_{1}^2 t^2} \,,
\end{equation}
which is singular at $t=0$. The Einstein tensor is given by
\begin{equation}
G_{\mu\nu}
=
\begin{bmatrix}
   \frac{4c_{1}^2 (m +mn +n)-3\alpha^2 }{4c_{1}^2 t^2} & 0 & 0 & -\frac{\alpha (m+n-2) }{2t} \\
    0 & c_{1}^2 ( (1-n)(m+n)-m^2)+\alpha^2/4  & 0   & 0  \\
    0 & 0 &  G_{22} & 0 \\
 -\frac{\alpha (m+n-2) }{2t} & 0 & 0   & G_{33}
\end{bmatrix},
\end{equation}
where \begin{equation}
G_{22}=\frac{(\alpha^2  - 4m^2 c_{1}^2 )c_{2}^2 t^{2(n-1)} e^{\alpha x} }{4 c_{1}^2}, ~ G_{33}=\frac{(\alpha^2  - 4n^2 c_{1}^2 )c_{3}^2 t^{2(m-1)} e^{\alpha x} }{4 c_{1}^2}\,.
\end{equation}
For a realistic cosmological model, we impose positive energy condition on the dynamical energy density
\begin{equation}
 \rho(t) = \frac{4c_{1}^2 (m +mn +n)-3\alpha^2 }{4c_{1}^2 t^2} \,,
\end{equation}
which is positive if $4c_{1}^2 (m +mn +n)-3\alpha^2 \geq 0,$ therefore we obtain a critical bound on the evolution factor $\alpha$
\begin{equation}
 \left|\alpha \right| \leq 2|c_{1}|\sqrt{\frac{m+mn+n}{3}} \,.
\end{equation}
Since the quantity in the square-root must be positive therefore we obtain an extra condition on the powers $m$ and $n$, i.e., $m> -n/(n+1), ~ n\neq -1$. If $n$ is a positive number then there are two possibilities that either $m$ is positive or negative. If it is negative then the spacetime becomes singular at $t=0$, while it is non-singular in the other case. For $-1<n<0$, the evolutionary function $B(t)=c_{1}t^m,$ has a positive power. However in the other case $n<-1$, the evolutionary function can have a positive or negative power as before. Therefore the evolution of above spacetime is such that it started at an initial time $t=c_{4}\neq 0$, then the energy density continue to decrease and vanishes as $t\rightarrow \infty$. The flux across $x$ and $y$ surfaces is zero therefore the density of first two components of linear momentum is zero however the $z-$ component of linear momentum density is
\begin{equation}
p(t)= G_{03}=  -\frac{\alpha (m+n-2) }{2t} \,,
\end{equation}
which could be positive or negative depending on the choice of $\alpha$, $m$ and $n$ and it asymptotically decays as $t\rightarrow \infty$.

Similarly in the other case the spacetime has the form
\begin{align}
   \textbf{2b.} \quad \mbox{d}s^2 = \mbox{d}t^2 - c_{1}^2 \mbox{d}x^2 -  e^{\alpha x}  ( c_{2}^2 e^{2\beta t}\mbox{d}y^2 +c_{3}^2 e^{2\gamma t} \mbox{d}z^2), \quad \beta \neq \gamma \,,
 \end{align}
which also has non-zero components of the Weyl tensor therefore it is not conformally flat. Again there are eleven non-zero curvature invariants and the Ricci scalar is
\begin{equation}
I^{\mathcal{R}}_{1,b}=\frac{3\alpha^2 - 4c_{1}^2 (\beta^2 +\beta \gamma + \gamma^2 )}{2c_{1}^2} \,,
\end{equation}
which is non-singular and non-dynamical unlike the previous case. The matter tensor is given by
\begin{equation}
G_{\mu\nu}
=
\begin{bmatrix}
   \frac{4\beta \gamma c_{1}^2 -3\alpha ^2}{4c_{1}^2} & 0 & 0 & -\frac{\alpha (\beta +\gamma ) }{2} \\
    0 & \frac{\alpha ^2 }{4} -c_{1}^2 ( \gamma^2 +\beta \gamma +\beta^2) & 0   & 0  \\
    0 & 0 &  \frac{(\alpha^2  - 4\gamma^2 c_{1}^2 )c_{2}^2  e^{\alpha x+2\beta t} }{4 c_{1}^2}& 0 \\
 -\frac{\alpha (\beta +\gamma ) }{2} & 0 & 0   &  \frac{(\alpha^2  - 4\beta^2 c_{1}^2 )c_{3}^2  e^{\alpha x+2\gamma t} }{4 c_{1}^2}
\end{bmatrix},
\end{equation}
In order to obtain a bound on the coefficient $\alpha$, we consider the density
\begin{equation}
 \rho(t) = \frac{4\beta \gamma c_{1}^2 -3\alpha ^2}{4c_{1}^2} \,,
\end{equation}
which is positive if
\begin{equation}
 |\alpha| \leq 2 |c_{1}|\sqrt{\frac{\beta \gamma }{3}} \,.
\end{equation}
Now there are two cases that either $\beta $ and $\gamma $ are both positive or negative. Unlike the previous case the above spacetime has a fixed energy density for all time. On the other hand the momentum density is also fixed which could be positive if both $\beta$ and $\gamma$ are positive and negative otherwise.

 \textbf{Case 3.} (6-dimensional algebra) \newline
The spacetime is
 \begin{align}
 \textbf{3a.} \quad   \mbox{d}s^2 = \mbox{d}t^2 - c_{1}^2 t^2 \mbox{d}x^2 -  e^{\alpha x}  ( c_{2}^2 t^{2n} \mbox{d}y^2 + c_{3}^2 t^{2n}\mbox{d}z^2),
 \end{align}
where the Weyl tensor vanishes therefore it is conformally flat and Petrov type O. There are four curvature invariants where the Ricci scalar is
\begin{equation}
I^{\mathcal{R}}_{1,a}=  \frac{ 3(\alpha^2 - 4n^2 c_{1}^2)}{2c_{1}^2 t^2}.
 \end{equation}
The matter tensor becomes
\begin{equation}
G_{\mu\nu}
=
\begin{bmatrix}
  \frac{4n(n+2)c_{1}^2-3\alpha^2}{4c_{1}^2 t^{2}} & 0 & 0 & \frac{(1-n)\alpha}{t} \\
    0 &  \frac{\alpha^2}{4}+nc_{1}^2 (2-3n) & 0   & 0  \\
    0 & 0 & \frac{(\alpha^2 -4n^2c_{1}^2)c_{2}^2t^{2(n-1)}e^{\alpha x}}{4c_{1}^2 }	& 0 \\
  \frac{(1-n)\alpha}{t} & 0 & 0   & \frac{(\alpha^2 -4n^2c_{1}^2)c_{3}^2t^{2(n-1)}e^{\alpha x}}{4c_{1}^2 }
\end{bmatrix},
\end{equation}
therefore the energy density is positive $G_{00}=\rho(t) >0$, for the following critical bound on $\alpha$
\begin{equation}
|\alpha | < 2|c_{1}| \sqrt {\frac{n(n+2)}{3}}\,,
\end{equation}
where the above spacetime is realistic. The quantity $n(n+2)$ must be non-negative which is true if $n>0$ or $n<-2$. The energy density vanishes as $t\rightarrow \infty$, so is the momentum density for the above spacetime.

In the other subcase the spacetime is
\begin{align}
  \textbf{3b.} \quad  \mbox{d}s^2 = \mbox{d}t^2 - (c_{1}t+c_{2})^2 \mbox{d}x^2 -  e^{\alpha x}  ( c_{3}^2 \mbox{d}y^2 + c_{4}^2\mbox{d}z^2),
 \end{align}
where the Weyl tensor vanishes therefore it is conformally flat and Petrov type O. The Einstein tensor becomes
\begin{equation}
G_{\mu\nu}
=
\begin{bmatrix}
 \frac{-3\alpha ^2 }{4(c_{1}t+c_{2})^2} & 0 & 0 & \frac{\alpha c_{1}}{c_{1}t+c_{2}} \\
    0 &  \frac{\alpha^2}{4}  & 0   & 0  \\
    0 & 0 & \frac{c_{3}^2 \alpha ^2 e^{\alpha x}}{4(c_{1}t+c_{2})^2 }	& 0 \\
 \frac{\alpha c_{1}}{c_{1}t+c_{2}} & 0 & 0   & \frac{c_{4}^2 \alpha ^2 e^{\alpha x}}{4(c_{1}t+c_{2})^2 }	
\end{bmatrix},
\end{equation}
therefore the energy density is negative $G_{00}=\rho <0$, for all time and the above spacetime is unrealistic.

\textbf{Case 4.} (7-dimensional algebra)\newline
  In this case the spacetime involve one arbitrary function
 \begin{align}
  \mbox{d}s^2 = \mbox{d}t^2 - A(t)^2\left ( \mbox{d}x^2 - e^{\alpha x}  ( c_{1}^2\mbox{d}y^2 + c_{2}^2\mbox{d}z^2) \right ),
 \end{align}
which turns out to be the case where Weyl tensor is zero thus the above spacetime is conformally flat and is of the Petrov type O. This is an interesting case as it can be regarded as an inhomeogeneuos extension of the FRW spactime. There are four curvature invariants including the basic Ricci scalar
\begin{align}
&I^{\mathcal{R}}_{1}= \frac{3(\alpha^2 - 4 A A^{\prime \prime} - 4 A^{\prime 2})}{2A^2}, \quad I^{\mathcal{R}}_{2} = \frac{3(4 A A^{\prime \prime} - 4 A^{\prime 2} + \alpha^2)^2}{64 A^4}\, \nonumber ,\\
&I^{\mathcal{R}}_{3}= \frac{3(4 A A^{\prime \prime} - 4 A^{\prime 2} + \alpha^2)^3}{512 A^6}, \quad
I^{\mathcal{R}}_{4}= \frac{21(4 A A^{\prime \prime} - 4 A^{\prime 2} + \alpha^2)^4}{16384 A^8} \, , \label{ci}
\end{align}
The Einstein tensor assumes the form
\begin{equation}
G_{\mu\nu}
=
\begin{bmatrix}
  \frac{3(4A^{\prime 2}-\alpha^2) }{4A^2} & 0 & 0 & 0 \\
    0 & \frac{(\alpha^2 - 8 A A^{\prime \prime} -4 A^{\prime 2})}{4} & 0   & 0  \\
    0 & 0 & \frac{c_{1}^2(\alpha^2 - 8 A A^{\prime \prime} -4 A^{\prime 2})e^{\alpha x}}{4}  & 0 \\
   0 & 0 & 0   &  \frac{c_{2}^2(\alpha^2 - 8 A A^{\prime \prime} -4 A^{\prime 2})e^{\alpha x}}{4}
\end{bmatrix} \label{E6},
\end{equation}
where there is no non-zero component of the linear momentum density. The energy density
\begin{equation}
\rho(t) = \frac{3(4A^{\prime 2}-\alpha^2) }{4A^2}\,,
\end{equation}
which could be positive or negative, in general. We now consider two dynamical evolutions of this non-flat spacetime in terms of a power-law and an exponential law. In case of a power-law $A(t)=c_{1}t^{m}$, the density $\rho(t)\rightarrow \infty$ for $m<0$ and  $\rho(t)\rightarrow 0$ for $m>0$, as $t\rightarrow \infty$, where the former is not an interesting case while the later contains some physical information. For example the density takes the form
\begin{equation}
\rho(t) = 3m^2t^{-2} - \frac{3\alpha^2}{4c_{1}^2} t^{-2m}\,,
\end{equation}
and its behavior could be examined by choosing an initial point
\begin{equation}
t_{0}= \left( \frac{2m c_{1}}{\alpha }\right) ^{1/(1-m)}\,, \quad m\neq 1
\end{equation}
where it is zero and it was negative before. Therefore we obtain three different cases (i) $0<m< 1$ (ii) $m= 1$ and (iii) $m>1$.

In the first case we choose $m=1/2$, without loss of generality and obtain
\begin{equation}
\rho(t) = \frac{3}{4} \left (1 - \frac{\alpha^2 t}{c_{1}^2}\right ) t^{-2},
\end{equation}
which is zero at $t=c_{1}^2/\alpha^2 > 0$. The critical point of $\rho(t)$ is $t_{c}=2c_{1}^2/\alpha^2$, where the density attains a global minimum because $\rho^{\prime \prime}(t_{c}) >0$. Since $\rho(t_{c})<0$, is negative therefore it is not a realistic spacetime. For the second case $m= 1$, the behavior of energy density is such that it starts from an initial value and decreases continuously till it vanishes and the global maximum is the initial point where the energy density has started at some non-zero time. The graph of energy density is depicted in Figure 1. In the last case $m>1$, it attains a global maximum value at
\begin{equation}
t_{\mbox{max}}= \left (\frac{4m c_{1}^2}{\alpha^2   } \right) ^{1/2(1-m)}, \quad m\neq 1
\end{equation}
after which it continue to decrease till it vanishes as $t\rightarrow \infty$, as is shown in Figure 2. By introducing a Hubble type parameter responsible for the possible expansion or contraction of this spacetime we find that
\begin{equation}
H(t)= \frac{A^{\prime}}{A}= \frac{m}{t} \,,
\end{equation}
therefore $H(t) \propto t^{-1}$, and this model mimics the behavior of a flat matter dominated universe (Einstein-de Sitter universe). Note that  in which the expansion function carry the same form for $m=2/3$, in the above case. It is interesting that our model is not flat and carries an energy density which is different from the energy density in a flat Einstein-de Sitter universe which varies as $\rho_{m}(t) \propto t^{-2}$. In our non-flat spacetime the energy density varies $\rho(t)\propto (4/3-3\alpha^2t^{2/3}/4c_{1}^2) t^{-2}$ in the case $m=2/3$, where the convergence of $\rho(t)$ is faster in comparison to $\rho_{m}$. In the case of radiation dominated universe the energy density also varies in the same way. However it is constant for a flat spacetime dominated only by vacuum energy. Note that in the case of radiation dominated flat universe the expansion factor is proportional to $t^{1/2}$, that lies in our first case. The energy density varies in the same way $\rho_{\gamma}\propto t^{-2}$, which in the above non-flat spacetime is $\rho(t)\propto (1-\alpha^2t/c_{1}^2) t^{-2}$.
\begin{figure}[ht]
\centering
\begin{minipage}[b]{0.45\linewidth}
\includegraphics[width=2.5in]{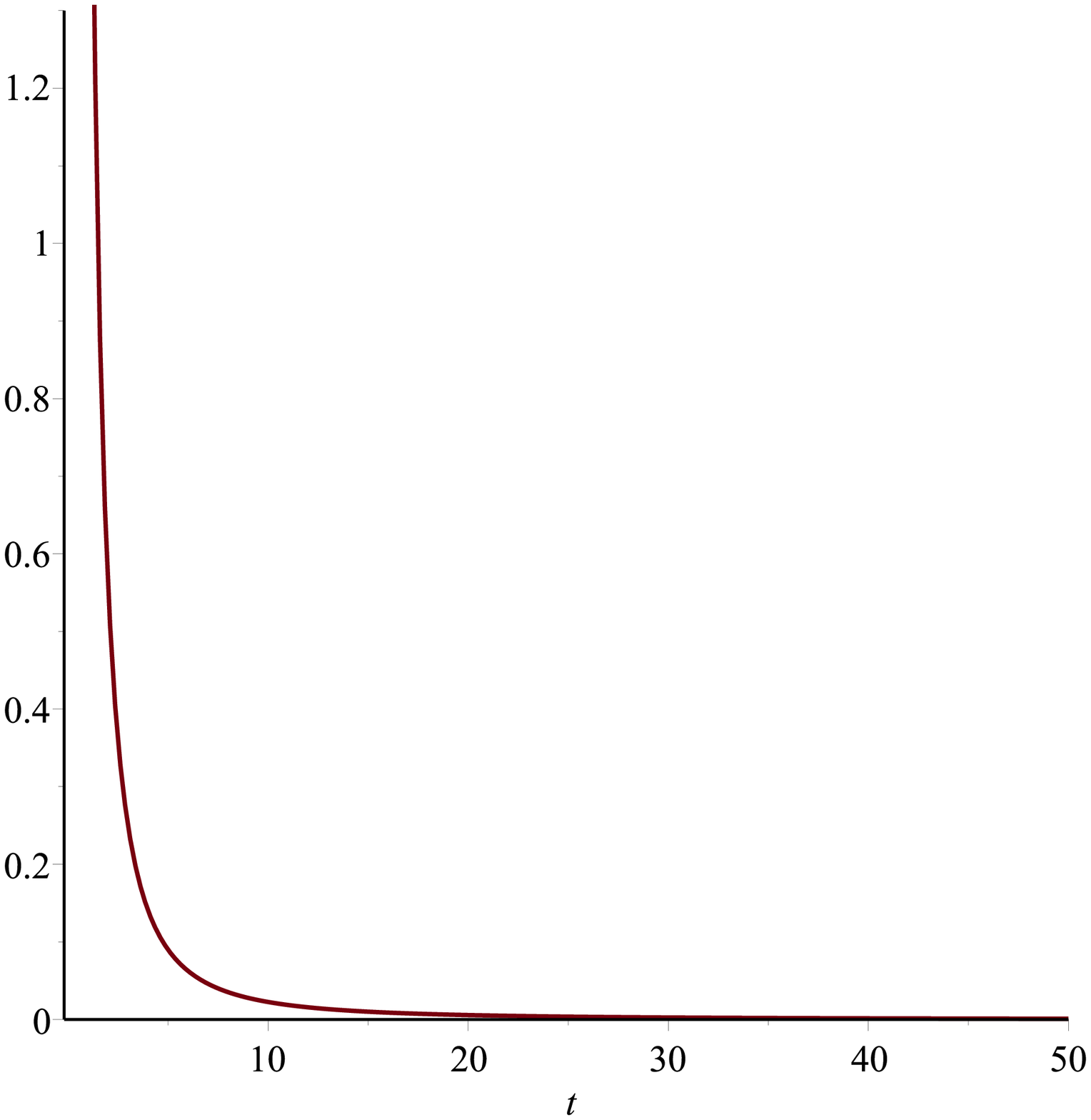}
\caption{The graph of energy density for $\alpha=1,m =1 , c_{1}=1$}
\end{minipage}
\begin{minipage}[b]{0.45\linewidth}
\includegraphics[width=2.5in]{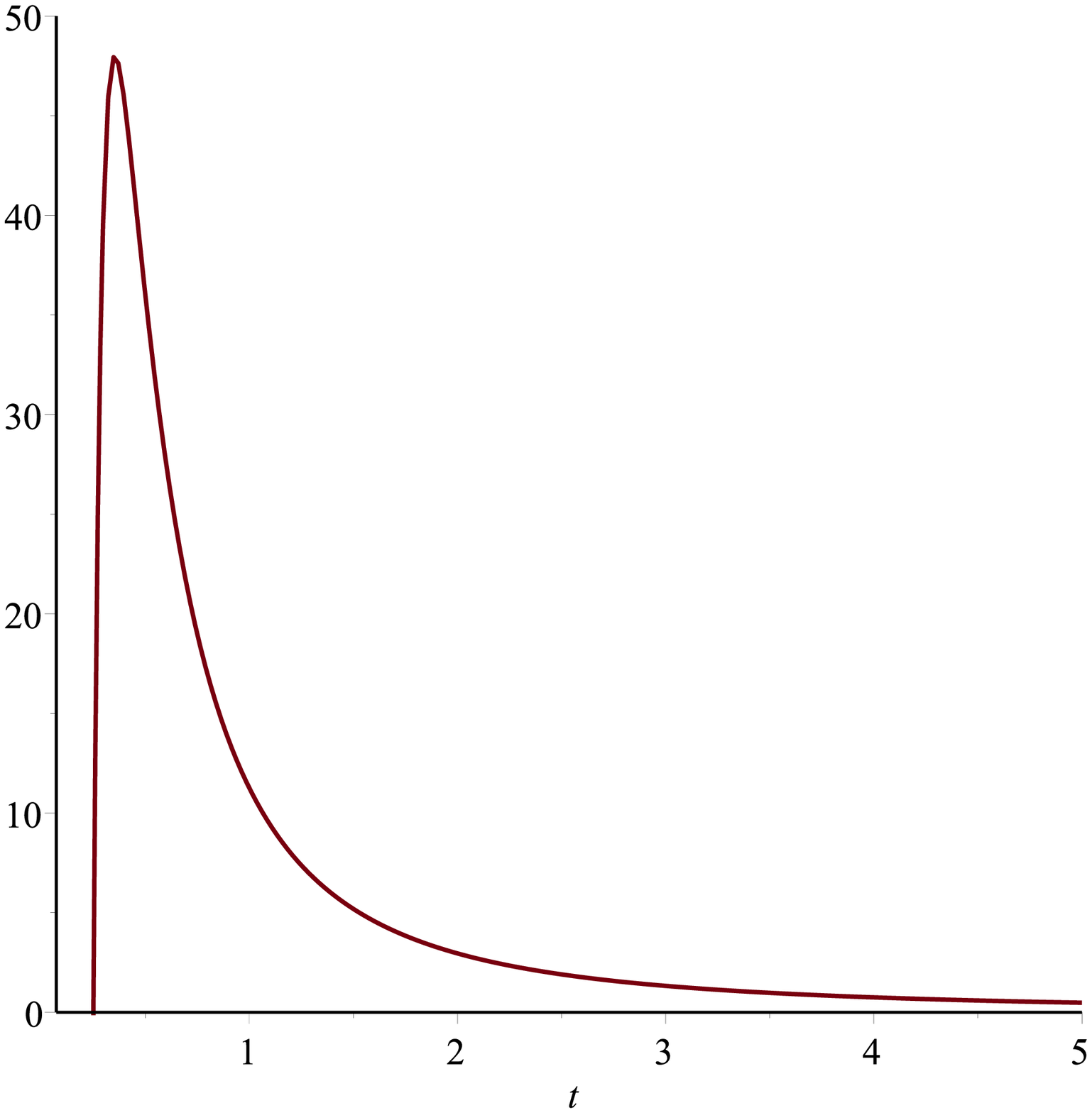}
\caption{The graph of energy density for $\alpha=1,m =2 , c_{1}=1$}
\end{minipage}
\end{figure}

In the case of an exponential form $A(t)= c_{1}e^{\beta t}$, the energy density becomes
\begin{equation}
\rho(t) = \frac{3\beta^2}{4}- \frac{\alpha^2}{4c_{1}^2} e^{-2\beta t}\,,
\end{equation}
which has the positive initial value $\rho(0)\geq 0$, if $\alpha \leq \sqrt{3} c_{1}\beta$. The case in which $\beta<0$, represents a contracting non-flat spacetime that carries a positive energy for a short time and continue to decrease such that $\rho \rightarrow -\infty$ as $t\rightarrow \infty$, because the contribution of the dynamical term will dominate the first term at some point. On the other hand an expanding non-flat spacetime requires that $\beta >0$, in which case we observe that the energy density increases rapidly for a short time where after it remains constant such that at an asymptotic limit $t\rightarrow \infty$,
it attains a finite positive value $3\beta^2/4$. The graph of energy density for this case is given in Figure 3. It is unexpected because in an expanding flat spacetime we expect that the energy density decreases indefinitely, however the curvature in this spacetime confine it to attain a positive definite value at an asymptotic limit. The Hubble parameter for this spacetime is non-dynamical $H(t)=\beta >0$, therefore this non-flat spacetime mimics a flat universe dominated by vacuum energy which has fixed density. However this non-flat spacetime has a varying energy density such that it attains a fixed positive value after a short interval of time.
\begin{figure}[ht]
\centering
\begin{minipage}[b]{0.45\linewidth}
\includegraphics[width=2.5in]{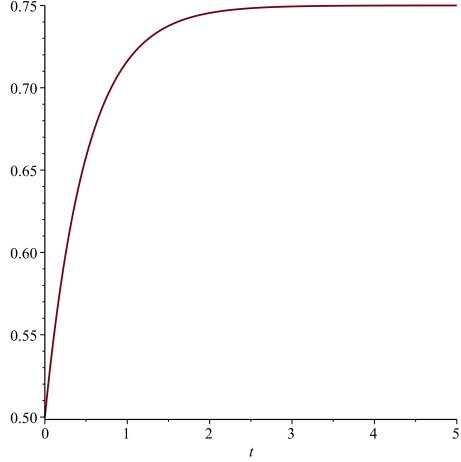}
\caption{The graph of energy density for $\alpha=1,\beta =1 , c_{1}=1$}
\end{minipage}
\end{figure}

 \textbf{Case 5.} (9-dimensional algebra)\newline
Following subcases arise in which we obtain nine-dimensional Noether algebra. However the form of Einstein tensor can easily be obtained from the last case by substituting the value of arbitrary function $A(t)$, in (\ref{E6}).
 \begin{align*}
 &\textbf{5(a)}\quad  \mbox{d}s^2 = \mbox{d}t^2 - c_{1}^2 \mbox{d}x^2 -  e^{\alpha x}  ( c_{2}^2\mbox{d}y^2 + c_{3}^2\mbox{d}z^2), \\
 &\textbf{5(b)} \quad
  \mbox{d}s^2 = \mbox{d}t^2 - (c_{1}t+c_{2})^2( \mbox{d}x^2 + e^{\alpha x}  ( \mbox{d}y^2 + \mbox{d}z^2)),
 \\
  &\textbf{5(c)}
\quad  \mbox{d}s^2 = \mbox{d}t^2 - c_{1}^2 \mbox{d}x^2 - e^{\alpha x + \beta t}  ( c_{2}^2\mbox{d}y^2 +c_{3}^2\mbox{d}z^2).
  \end{align*}
The curvature rank for the spacetimes in case (5b) is 3 and the Lie algebra of both curvature and Weyl collineations is infinite dimensional \cite{cam}. In the first case we obtain following non-zero components of the Einstein tensor
\begin{equation}
G_{\mu\nu}
=
\begin{bmatrix}
 -\frac{3\alpha ^2}{4c_{1}^2}  & 0 & 0 & 0 \\
    0 & \frac{\alpha ^2}{4}& 0   & 0  \\
    0 & 0 &\frac{\alpha ^2 c_{2}^2 e^{\alpha x}}{4c_{1}^2}  & 0 \\
   0 & 0 & 0   &   \frac{\alpha ^2 c_{3}^2 e^{\alpha x}}{4c_{1}^2}  
\end{bmatrix} .
\end{equation}
Since the energy density is negative therefore the above spacetime is unrealistic although it is conformally flat. The components of the matter tensor in case (5b) can be obtained directly by placing the value of $A(t)=c_{1}t+c_{2},$ in case 4, equivalently
\begin{equation}
G_{\mu\nu}
=
\begin{bmatrix}
\frac{3(4c_{1}^2-\alpha ^2)}{2(c_{1}t+c_{2})^2} & 0 & 0 & 0 \\
    0 &  \frac{\alpha ^2-4c_{1}^2}{4} & 0   & 0  \\
    0 & 0 &  \frac{(\alpha ^2-4c_{1}^2)e^{\alpha x}}{4}   & 0 \\
   0 & 0 & 0   &  \frac{(\alpha ^2-4c_{1}^2)e^{\alpha x}}{4}
\end{bmatrix} .
\end{equation}
The positive energy condition requires that $|\alpha| <2|c_{1}|$, which indicates that the expansion coefficient $\alpha$ must be smaller than the twice the slope of graph of $A(t)=c_{1}t+c_{2}$. The energy density vanishes as $t\rightarrow \infty$.

Lastly the case (5c) has a distinct difference from the previous two cases that there arise non-zero off-diagonal terms in the matter tensor. Following are the non-zero components of the Einstein tensor
\begin{equation}
G_{\mu\nu}
=
\begin{bmatrix}
-\frac{(3\alpha ^2-4\beta^2 c_{1}^2)}{4c_{1}^2} & 0 & 0 & -\alpha \beta \\
    0 &  \frac{\alpha ^2-12\beta ^2 c_{1}^2}{4} & 0   & 0  \\
    0 & 0 &   \frac{(\alpha ^2-4\beta ^2 c_{1}^2)c_{2}^2~e^{\alpha x + 2\beta t }}{4c_{1}^2}  & 0 \\
  -\alpha \beta & 0 & 0   &        \frac{(\alpha ^2-4\beta ^2 c_{1}^2)c_{3}^2~e^{\alpha x+ 2\beta t}}{4c_{1}^2} 
\end{bmatrix},
\end{equation}
where we must have $|\alpha| < 2/\sqrt{3}\,|c_{1}\beta|$, for a realistic model. Note that the energy density in this Bianchi V spacetime is fixed and it attains non-zero components of the momentum density unlike the previous cases where these were zero.

   \textbf{Case 6.} (10-dimensional algebra)\newline
There are two subcases of 10-dimensional Noether algebra which are
 \begin{align*}
 &\textbf{6(a)}\quad  \mbox{d}s^2 = \mbox{d}t^2 - c_{1}^2 \mbox{d}x^2 - e^{2\beta(c_{1}x+t )} ( c_{2}^2\mbox{d}y^2 + c_{3}^2\mbox{d}z^2), \\
 &\textbf{6(b)} \quad
  \mbox{d}s^2 = \mbox{d}t^2 - c_{1}^2 \mbox{d}x^2 - e^{2\beta(-c_{1}x+t )} ( c_{2}^2\mbox{d}y^2 + c_{2}^2\mbox{d}z^2).
   \end{align*}
Although the above models are conformally flat but these are unrealistic spacetimes as the Einstein tensor is
 \begin{equation}
G_{\mu\nu}=
\begin{bmatrix}
  -2 \beta^{2} & 0 & 0 & \mp 2c_{1} \beta^{2} \\
    0 &-2c_{1}^2 \beta^{2} & 0   & 0  \\
    0 & 0 & 0  & 0 \\
   \mp 2c_{1} \beta^{2} & 0 & 0   &  0
\end{bmatrix},
\end{equation}
and the energy density in both sub cases is negative, $G_{00}=- 2\beta^2 <0$.

\section{Some Considerations into $f(R)-$Gravity Theory}
We now develop a few considerations for the anisotropic cosmological solutions that are obtained as a consequence of the existence of Noether symmetries in the context of $f(R)-$gravity. The idea is to probe into earlier epochs in the evolution of the Universe using the results obtained in the previous section. As there arises several cases therefore we confine ourself to a few most relevant cases for our purpose. For example let us consider the case (5b) which meets the positive energy condition for ($|\alpha|<2|c_{1}|)$ given by
\begin{equation}
\mbox{d}s^2 = \mbox{d}t^2 - (c_{1}t+c_{2})^2( \mbox{d}x^2 + e^{\alpha x}  ( \mbox{d}y^2 + \mbox{d}z^2)), \label{fmetric}
\end{equation}
in which case the Ricci scalar is
\begin{equation}
R = \frac{3(\alpha^2-4c_{1}^2)}{2(c_{1}t+c_{2})^2}\,, \label{ricci}
\end{equation}
which remains negative for all time due to positive energy condition therefore the underlying manifold is hyperbolic. We now obtain the field equations using the action of $f(R)-$gravity involving a matter term in standard gravitational units \cite{cap3,sot}
\begin{equation}
\mathcal{A} = \frac{1}{2\kappa} \int 	\mbox{d}^4 x \sqrt{-g} \, f(R) + S_{M}(g_{\mu\nu},\psi) , \label{feqs}
\end{equation}
where $\kappa = 1$, $g$ denotes the background metric and $f(R)$ is a general function of first curvature invariant which in our case is given by (\ref{ricci}). Here $\psi$ denotes all matter fields coupled to gravity. The field equations now assume the form 
\begin{equation}
f^{\prime} R_{\mu\nu} - \frac{1}{2} f(R) g_{\mu\nu} - \Big (\nabla_{\mu}\nabla_{\nu}- g_{\mu\nu}\Box \Big) f^{\prime} =  T_{\mu\nu} \,,  
\end{equation}
where $f^{\prime} = df(R)/dR$, $\nabla_{\mu}$ is the covariant derivative with respect to the Levi-Civita connection of the metric and $\Box = \nabla^{\mu}\nabla_{\mu}$. An immediate consequence of the above field equations is that their trace yield an important equation 
\begin{equation}
Rf^{\prime} - 2f(R) + 3 \,\Box (f^{\prime}) = T, \label{const}
\end{equation}
which indicates how matter part $T$, in $f(R)$-gravity is differentially connected to the curvature in the given spacetime unlike in standard GR, where we have an algebraic relationship $R=-\kappa T$. We now employ Bianchi V spacetime (\ref{fmetric}) and solve the fields equations in a relatively simple case of a dust cloud. As in our case the Ricci scalar $R$ is a function of $t$, therefore the field equations yield a system of ordinary differential equations $f^{\prime}(R(t))= \mathcal{F}(t)$, where we subsequently represent the time derivative with an overdot (not to be confused with an overdot in Section 1). In this case the non-zero components of Ricci tensor are 
\begin{equation}
R_{11}=\frac{4c_{1}^2 -\alpha^2}{2}, ~R_{22} =R_{33}=e^{\alpha x} R_{11}\,,
\end{equation}
which upon using the field equations (\ref{feqs}) result into 
\begin{align} 
&  - \frac{1}{2}f(R) = T_{00} ,\label{seq1}\\
&(4c_{1}^2 -\alpha^2) \mathcal{F} - (c_{1}t+c_{2})^2 ( 2\ddot{\mathcal{F}}- f(R) ) = T_{11}. \label{seq2} 
\end{align}
We now assume a perfect fluid whose energy-momentum tensor is given by 
\begin{equation} 
T_{\mu\nu} = (\rho + p) u_{\mu} u_{\nu} - p g_{\mu\nu}, 
\end{equation}
where $\rho(t)$ and $p(t)$ are energy and pressure densities of the fluid which satisfy the equation of state 
\begin{equation}
p = \omega \rho, \quad 0 \leq \omega \leq 1 , 
\end{equation}
while $u_{\mu} = \sqrt{g_{00}}( 1,0,0,0)$, is the four velocity in comoving coordinates. The conservation equation yields 
\begin{equation} 
\dot{\rho} + (\rho +p) \left ( \frac{\dot{A}}{A}+ \frac{\dot{B}}{B}+\frac{\dot{C}}{C} \right) = 0, 
\end{equation}
which for the underlying spacetime (\ref{fmetric}) and in case of a pressure-less fluid $\omega =0$, assumes the form 
\begin{equation} 
\dot{\rho} + \frac{ 3c_{1}\rho}{c_{1}t+c_{2}}   = 0, 
\end{equation}
whose solution involving a constant of integration $C_{1}$ is 
\begin{equation} 
\rho(t) = \frac{C_{1}}{(c_{1}t+c_{2})^3} \,\,.
\end{equation}
By substituting the above value in (\ref{seq1}) and (\ref{seq2}), we obtain an ordinary differential equation  $\mathcal{F}(t),$ given by 
\begin{align} 
\ddot{\mathcal{F}} - \frac{(4c_{1}^2 - \alpha^2)}{2(c_{1}t+c_{2})^2}\mathcal{F} + \frac{C_{1}}{(c_{1}t+c_{2})^3} = 0, 
\end{align}
which has a solution 
\begin{align}
\mathcal{F} (t) = C_{2} (c_{1}t+c_{2} ) ^{\lambda_{+}}+ C_{3} (c_{1}t+c_{2} ) ^{\lambda_{-}} - \frac{2C_{1}}{\alpha^2 (c_{1}t+c_{2})} \,\,,
\end{align}
where $C_{2}$ and $C_{3}$ are constants of integration. The exponents $\lambda_{+}$ and $\lambda_{-}$, are given by 
\begin{equation} 
\lambda_{+} = \frac{c_{1} + \sqrt{9c_{1}^2 -2\alpha^2}}{2c_{1}}\,, \quad \lambda_{-} =  \frac{c_{1} - \sqrt{9c_{1}^2 -2\alpha^2}}{2c_{1}}\,,
\end{equation}
which are both real for positive energy condition. In order to find the explicit form of $f(R)$, we employ the equation (\ref{const}) to obtain 
\begin{equation}
f = -\frac{2C_{1}}{(c_{1}t+c_{2})^3}\,, 
\end{equation}
which in terms of $R$, using (\ref{ricci}) is given by 
\begin{equation}
f(R) \propto R^{3/2}.
\end{equation}
This is quite unexpected because the above relationship is found for $f(R)-$gravity in the background of FRW universe and it is observed that all other models other than this remain obscure \cite{cap3}. Therefore, an analysis of a dust cloud in the background of an \emph{anisotropic} and homogeneous Bianchi V spacetime (\ref{fmetric}) which does not correspond to an FRW universe, reveal the same form of $f(R)$. Here we have demonstrated a successful implementation of our results obtained in the previous section in $f(R)-$gravity. It would be interesting to investigate more physical models with different equations of state in the background of other Bianchi V spacetimes following the same lines. A detailed analysis of rest of the cases of such spacetimes and a brief comparison with other extended theories of gravity will be discussed elsewhere.

\section{Summary}
We used Noether symmetries to study Bianchi V cosmologies. It is identified that the algebra of Lie point symmetries of the geodesic equations in such spacetimes can have dimensions $5-8$, $10,12$ or $13$. On the other hand the dimensions of Noether algebras is $4-7$, $9$ or $10$. The presence of Noether symmetries helps us to obtain first integrals or constant of motion using Noether theorem corresponding to the geodesic Lagrangian. A key advantage of this approach is the determination of unknown functions in the Lagrangian which are specified while solving the determining equations for Noether symmetries that reduces the dynamics significantly. In all cases $2-6$, we explicitly specified the unknown evolutionary functions in these spacetimes. Besides one can use first integrals to obtain closed form of exact solutions of the equations of motion, i.e., geodesics in non-flat spacetimes.

In order to interpret our results physically we used positive energy condition which holds for all realistic cosmological models. Interestingly it imposes constraints on the solutions and provide us critical bounds on the constants for which a Bianchi V spacetime is physical. The expansion coefficient $\alpha$, in all Bianchi V spacetimes is crucial. By imposing the positive energy condition on the obtained solutions it turns out that we can investigate different possible critical bounds on $\alpha$, in which the non-flat spacetimes carry interesting physical characteristics. The positive energy condition further imposes constraints on the spacetimes and we obtain cases in which the energy density behaves in one of the following ways. It is positive and constant for all time. It varies with time and attains a global maximum after sometime where after it asymptotically converges to a relatively smaller but positive value. It increases for all time and attains a maximum value at the asymptotic limit $t\rightarrow \infty$. A brief comparison of flat models (vacuum, radiation, matter) with the non-flat spacetimes is given in detail.

Another important consequence of this study is the identification of $f(R) \propto R^{3/2}$, for a dust cloud in an anisotropic and homogeneous spacetime (\ref{fmetric}) which is proven to exist in FRW universe using standard $f(R)-$gravity approach. It indicates us to briefly probe into earlier epochs of our Universe when it was least isotropic. Therefore those Bianchi V spacetimes where the positive energy condition successfully holds are the best candidate for further investigation with more physical models with different equations of state.

\section*{Acknowledgments} IH would like to thank TWAS-UNESCO for awarding Associateship
at Kavli Institute for Theoretical Physics, Chinese Academy of Sciences, Beijing, China
where this work was finalized. We would like to thank the referee for suggesting us to develop considerations for $f(R)-$gravity in the light of obtained results.

%%%%%%%%%%%%%%%%%%%%%%%%%%%%%%%%%%%%%%%%%%%%%%%%%%%%%%%%%%%%%%%%%%%%%%%%%%%%%%%%%%%%%%%%%%%%%%%%%%%%%%%%%%%%%%%%%%%%%%%
%%%%%%%%%%%%%%%%%%%%%%%%%%%%%%%%%%%%%%%%%%%%%%%%%%%%%%%%%%%%%%%%%%%%%%%%%%%%%%%%%%%%%%%%%%%%%%%%%%%%%%%%%%%%%%%%%%%%%%%

\end{document}